\documentstyle[prl,aps]{revtex}
\input{psfig}
\draft
\begin{document}
\twocolumn[\hsize\textwidth\columnwidth\hsize\csname@twocolumnfalse\endcsname
\title{Evolution  on a Rugged Landscape:
Pinning and Aging}
\author{I. Aranson $^{1,2}$, L. Tsimring$^3$,  and V. Vinokur $^2$}
\address{
$^1$ Bar Ilan University, Ramat Gan 52900, Israel\\
$^2$ Argonne National Laboratory,
9700 South  Cass Avenue, Argonne, IL 60439 \\
$^3$Institute for Nonlinear Science, University of California, San Diego, La Jolla, 
CA 92093-0402
}
\date{\today}
\maketitle
\begin{abstract}
Population dynamics 
on a rugged landscape is studied analytically and numerically within a
simple discrete model for evolution of $N$ individuals in one-dimensional
fitness space.  We reduce the set of master equations to a single
Fokker-Plank equation which allows us to describe the dynamics of the 
population in terms of thermo-activated Langevin diffusion  of a single particle 
in a specific random potential. We found that the randomness in the mutation rate 
leads to pinning of the population and on average to a logarithmic 
slowdown of the evolution, resembling aging phenomenon in spin glass systems.
In contrast, the randomness in the replication rate turns out to be 
irrelevant for evolution in the long-time limit as it is smoothed out by 
increasing ``evolution temperature".  The analytic results are in a good 
agreement with numerical simulations.
\end{abstract}
\pacs{PACS: 87.10.+e,82.20.Mj,64.60.Cn,05.40.+j}
\vskip1pc]
\narrowtext

Recently, a simple model was introduced to describe
the evolution of a finite population of mutating species in 
a one-dimensional fitness space\cite{1,2}. Every species is characterized
by a fitness variable which controls its replication 
rate, and mutations move the species equiprobably up or down along this 
fitness axis. Analysis of this model leads 
to a two-staged dynamics of a population initially spread within some 
fitness range. At the first (fast) stage, the population forms a universal 
pulse-like distribution in the fitness space, which usually is accompanied
by the fast growth of the mean fitness. At the second stage, the pulse 
propagates toward higher fitness due to mutations, new more fit mutants
are constantly generated, and less fit die due to the constant
population size constraint. The mean fitness grows linearly in time. 
The non-trivial scaling of the fitness growth with the mutation strength 
and the population size was found in Refs. \cite{1,2} and appears to be 
in a qualitative agreement with the evolution dynamics of RNA virus\cite{RNA}. 

Clearly, this model contains an implicit assumption that underlying
fitness landscape in the configurational space is smooth. 
In reality, fitness landscapes likely exhibit substantial degree of 
ruggedness\cite{kauf}, i.e. a probability to find a more (or less) 
fit mutant varies for different genomes.
Moreover, for rugged landscapes there may be genomes which are more fit
than any of their nearest neighbors in the sequence space which can be 
accessed as a result of a point mutation. Thus, there are local maxima 
of fitness. Rugged fitness landscapes are usually considered in 
multi-dimensional configurational spaces of individual species where the 
fitness is a complicated function of the detailed 
structure of the genome. A popular $NK$-model for the relation between 
genomic structure and fitness was introduced 
by Kauffman\cite{kauf}. Many other more complicated or realistic models of 
rugged landscapes are described in the literature (see, e.g., 
\cite{fontana,wolynes}). 

Our model of one-dimensional fitness space is essentially different 
from these models since individual species with different internal
structure but identical fitness are considered indistinguishable and 
therefore assigned to the same location along the fitness 
axis\cite{perelson}. This model obviously cannot 
account for the evolution of the internal structure in the population, in 
particular, the famous error catastrophe which occurs at high mutation rate 
and leads to accumulation of ``bad mutants" and perpetual loss of heritable 
genetic information\cite{eigen}. Nevertheless our model of evolution in a 
one-dimensional fitness space is capable of reproducing important dynamic 
features of evolution found in more realistic models.
Due to its relative simplicity, it is more amenable to analytical 
treatment then multi-dimensional configurational space models.

The goal of this paper is to study the dynamics of the population 
evolution in the framework of the model\cite{1,2} which however 
takes into account ruggedness of the underlying fitness landscape. 
Random fluctuations which are considered to be 
functions of one fitness variable characterizing the genotype as a whole, 
modify locally both replication rate as well as mutation probability. 
We assume that the replication rate is a sum of linear function 
of fitness and random fluctuations. Linear part provides 
the selective pressure driving the population towards higher fitness, 
as in smooth landscape case. Random fluctuations however provide trapping
of the population near local maxima of replication rate and on average 
slow down the mean fitness growth. Probability of mutation up and down in
fitness at a given state also has fluctuating part. We assume that these 
fluctuations are statistically independent of fluctuations of the 
replication rate. Based on master equations for underlying Markov
process, we derive the Fokker-Plank equation for probability distribution
of species in the population.  We find that initial growth of fitness
is dominated by the fluctuations of the replication rate, but at
long-time limit these fluctuations become irrelevant. The long-time 
limit is completely
determined by the fluctuations of the mutation rate which lead to
logarithmic slowdown of the fitness growth. This phenomenon is
analogous to the thermally-activated creep of a particle in quenched
random (pinning) potential\cite{blatter}.

Let us first specify the model (cf. Ref.\cite{2}). Consider a population 
of $N$ individuals which can replicate according to their replication 
rate $R$, and babies can mutate, thus changing their fitness relative 
to parents. We assume asexual replication, i.e. any individual has a 
chance to reproduce independently. Once a baby is born, some member 
of the whole population (including the new baby) which is picked  at 
random, is eliminated to preserve the constant population size. We 
assume that the replication rate $R$ is a function of only one independent 
variable $x$ (fitness), which has a linearly growing component and a random 
component, $R(x)=x+\xi_r(x)$ (without loss of generality we chose a unit
slope). When mutation occurs, a mutant baby changes its $x$ value to 
$x\pm 1$ with probabilities $M^{\pm}(x)=(1\pm\xi_\mu(x))\mu/2$,
respectively. Random functions $\xi_r$ and $\xi_\mu$ are assumed to be 
uniformly distributed between $\pm\Delta_{r,\mu}$ and they are ``frozen'',
or quenched, for a given realization. 

An exact description of this Markov process involves a very large 
(strictly speaking, infinite) set of master equations for the 
probabilities of all possible population configurations. Significant
simplification can be achieved in the limit of small mutation rate $\mu$
when all population is highly localized near $\langle x\rangle$ and only 
two neighboring sites $x$,$x+1$ are usually occupied simultaneously (the
probability to occupy  simultaneously 3 or more sites is smaller by a factor 
of $\mu$).
The number of  master equations then is reduced to $N$ (see \cite{2}).
It turns out that a much simpler system of only 2 individuals 
exhibits similar non-trivial features as $N$-individual model. This system 
in the limit of small $\mu$ can be described by only two equations for
probability $f$ to find a particle at site $x$, and $g$, the
probability of finding one particle at site $x$ and another at $x+1$,
\begin{eqnarray}
&&\partial_t f(x)  =  - \mu \frac{4}{3} R(x) f (x)
+\frac{1}{3} R(x) \left(g(x) + g(x-1) \right),  \label{eq0} \\
&&\partial_t g(x)  =  \frac{4}{3} \left(
R(x)M^+(x) f(x) + \right.     R(x+1)\times  \nonumber \\
&& \left. \times  M^-(x+1)f(x+1) \right) 
-\frac{1}{3} \left(R(x) + R(x+1) \right)g(x).  
\label{eq1}
\end{eqnarray}
The first term in r.h.s. of (\ref{eq0}) reflects the probability of
mutation events leading to transition from the collapsed state when 
both particles occupy site $x$ to either of distributed states 
$(x,x+1)$ or $(x-1,x)$. 
The second term describes a reverse process of replication at site $x$ 
which is followed by elimination of a particle at $x\pm1$ and collapse 
of the population to the site $x$. Terms in the right-hand side of 
Eq.(\ref{eq1}) have the same origin.  For $\mu\ll 1$, $\partial_t g(x)$ 
is small in the asymptotic limit, and $g$ is enslaved to $f$,
$$ g(x)=  4\frac{R(x)M^+(x) f(x) + R(x+1) M^-(x+1)f(x+1)} 
{R(x) + R(x+1)}$$.
Substituting $g(x)$ into Eq.(\ref{eq0}), taking a continuous
$x$ limit, and keeping only linear in $\xi$ terms, we obtain
a single equation for $f$,
\begin{eqnarray}
 \partial_t f(x)& =&  -\frac{1}{3} \mu \partial_x
\left[(1+2x\xi_\mu+ \partial_x \xi_r) f (x)\right]  \nonumber \\
&& +\frac{\mu}{3} \partial_x^2[(x+\xi_r
+\frac{1}{2}\xi_\mu) f(x)].
\label{eq3}
\end{eqnarray}
In a similar fashion but with more cumbersome algebra, 
an equation describing the probability 
distribution $f(x,t)$ for a population of $N$ individuals to be
localized at point $x$ at time $t$ can be derived in a
small $\mu$ limit. In the limit of $N \gg1$ it reads
\begin{eqnarray}
&&\partial_t f(x)= -\frac{\mu}{2} \partial_x
\left[(N+2 x\xi_\mu+N \partial_x \xi_r) f 
(x)\right]  \nonumber \\
&& +\frac{\mu}{2} \partial_x^2 [(x+\xi_r+ \frac{N}{2} \xi_\mu)
f(x)].
\label{eq4}
\end{eqnarray}
For $\Delta_{r,\mu}=0$ this equation coincides with one derived in
Ref.\cite{2} and describes the increase of average fitness 
of $N$ individuals as a result of random mutations and selective pressure. 
In the limit of smooth landscape the average
fitness $\langle x \rangle \equiv \int xf(x)dx$ grows linearly in time
with the rate $V_0 =\frac{\mu N}{2} $\cite{2}. 
A general solution to Eq.(\ref{eq3}) in the presence of random fluctuations 
is not available, however one can estimate $\langle x\rangle$ using general 
methods of stochastic kinetics in random media \cite{landau}.  To this end, 
note that Eq.(\ref{eq3}) has a form of Fokker-Plank equation for the dynamics 
of a single particle in the fitness space under the global bias $N\mu/2$  
in the presence of the random quenched potential. Note that in our case the 
temperature itself depends on $x$. The corresponding dynamic process is 
governed by the Langevin equation:
\begin{equation}
\frac{d x}{dt}=\frac{N}{2}(1+\partial_x \xi_r) + x\xi_\mu+\eta(x,t),
\label{langevin}
\end{equation}
where stochastic term $\eta(x,t)$ has the following correlator 
$\langle\eta(x,t)\eta(x,t^\prime)\rangle=$ $2 \tau \delta(t-t^\prime)$ with
local ``evolution temperature" $\tau=$ $\frac{1}{2}  (x+\xi_r+ N \xi_\mu/2 )$, 
and we rescaled time $\mu t\to t$. If the stochastic term were absent, the 
particle would have been  pinned by local minima of the quenched potential. 
However, in the presence of ``thermal fluctuations"  $\eta(x,t)$ the system 
will evolve via a sequence of thermally activated jumps from one minimum to 
a neighboring one favored by the global positive bias (selective pressure).
This type of motion is well known in the dynamics of disordered media and 
is usually referred to as creep\cite{blatter}.

We discuss the cases $\xi_\mu \ne 0, \xi_r = 0$ and  $\xi_\mu = 0, \xi_r \ne 0$
separately, since they exhibit different asymptotic behavior. 
Let us first focus on the case $\xi_\mu \ne 0, \xi_r = 0$. 
It generalizes a well-known Sinai diffusion problem for the particle
subject to a random force field. The particle displacement obeys ultra-slow 
sub-diffusion law $\langle x \rangle \sim (\log t)^2 $ for zero average 
driving \cite{sinai}, and the power-law dependence $\langle x \rangle \sim t^\kappa$  
in the driven case \cite{vinokur1}.   Our situation is more complicated since
the evolution temperature and the magnitude of disorder depend on
$x$ explicitly. To obtain the time-dependence of $\langle x \rangle$ 
we generalize the approach of Ref. \cite{vinokur1}. According to the 
general theory of stochastic growth\cite{landau} the evolution rate 
 $V=\langle \dot x\rangle\propto f(\bar{U}[x])$, where $\bar U$ is the typical 
potential barrier controlling the evolution, and $f$ is the corresponding 
quasi-equilibrium distribution function, given by (we consider the limit 
$x \gg N$ and therefore drop the random term in the expression for evolution 
temperature): 
\begin{equation} 
f(x) \sim \exp \left( 2 \int_0^x  \xi_\mu(x^\prime) d x^\prime + 
(N-1)\log(x) \right)
\label{distr}
\end{equation} 
Taking into account that the typical height of the potential 
barrier corresponding to the displacement $x$ is 
$\sqrt { \langle (\int_0^x  \xi_\mu(x^\prime) d x^\prime )^2 \rangle} $ $  = 
\Delta_\mu \sqrt{x /3} $ (by assumption $\xi_m$ is uniformly distributed 
between $\pm \Delta_\mu$) one finds the following estimate 
for the typical time to evolve over the distance $x$,
\begin{eqnarray} 
\langle t \rangle &\sim& V^{-1}\sim 1/f (\bar{U}[x])\nonumber\\
&\sim&\ \exp ( 2  \Delta_\mu\sqrt{x /3} - (N-1)\log x ).
\label{distr1} 
\end{eqnarray} 
Inverting Eq. (\ref{distr1}) we obtain the following evolution law 
\begin{equation}
\langle x \rangle \sim \frac{3}{4 \Delta_\mu^2 } \left( \log (t+ C_0) 
+ (N-1)\log \langle x \rangle \right)^2 
\label{distr2}
\end{equation} 
where  the  constant $C_0$ is determined by the initial condition for  
$\langle x\rangle$ at $t=0$.   In the limit $\langle x\rangle \to  \infty$, 
the final asymptotic behavior of Eq. (\ref{distr2}) coincides with the Sinai
diffusion law \cite{sinai}
$\langle x\rangle \sim \frac{1}{ \Delta_\mu^2 } ( \log  t)^2 $. 
For large $N$ however 
this regime realizes at {\it enormously} large times $t$ and distances 
$\langle x\rangle$ ($\langle x\rangle \gg N^2$ or  $ t\gg e^N$), 
and can be hardly observed in numerical experiment.
In the intermediate asymptotic regime ($1 \ll \langle x\rangle \sim N^2$), 
the last term in Eq. (\ref{distr2}) is a leading one, and taking into account 
that $\log \langle x\rangle$ is a slow function as compared to 
$\langle x \rangle$, we obtain the following evolution law:
\begin{equation}
\langle x\rangle \sim \frac{N} { \Delta_\mu^2 }  
\log (\mu t+C_0)   + x_0,
\label{distr3}
\end{equation}
(we restored original time units).  Constants  $C_0$ and $x_0$ 
themselves depend on $N$ and $\Delta_\mu$, and a crossover to disorder-free 
($\Delta_\mu \to 0$) linear growth of the fitness is recovered by taking 
the limit $ \mu t \ll C_0$ and $C_0 \to \infty$. 
Our numerical results for discrete model simulations confirm analytical 
formula (\ref{distr3}). Fig. 1 illustrates smooth logarithmic dependence 
$ \langle \langle x(t) \rangle \rangle $ which was obtained
numerically for $N=50,\ \mu=0.02, \Delta_\mu=0.7$ 
averaged over 50 statistically-independent configurations 
of quenched disorder (double angular brackets indicate computing a mean 
fitness of the entire population and
ensemble averaging over different realizations of $\xi$).  As expected, for 
large $t$ the rate of the evolution slows down which corresponds to the 
{\it aging phenomenon} known in glassy dynamics \cite{glass}. 
In the Fig. 1 we also show  the dynamics of the mean
fitness of the population $\langle x \rangle $ for a single run.
This dependence is characterized by long periods of relatively 
steady fitness 
level (stasis) interrupted by abrupt jumps of the fitness. 
During these long intervals of stasis the population is 
trapped in regions 
of low probability of escape towards larger $x$ (large negative $\xi_\mu$). 
Such an inhomogeneous pace of evolutionary changes is known in biology as a 
{\it punctuated equilibrium} hypothesis (see \cite{kauf}). 
We performed numerical 
experiments with our model for various $N, \mu$, and $\Delta_\mu$. Fitting 
the results to the logarithmic dependence $\langle \langle x  \rangle \rangle$
 $=A\log (t+t_0)+
x_0$ yields a good agreement with formula (\ref{distr3}). Figure 2 shows 
the dependences of $A\Delta_\mu^2$ on $N$ for several sets of parameters 
$N,\ \Delta_\mu$.   All the graphs collapse  close to a single straight line 
with slope 1 in log-log coordinates, as predicted by the theory. 

Now we turn to the case $\xi_\mu = 0$ but $ \xi_r \ne 0$ where particle 
dynamics is governed by the diffusion in the random potential 
$\xi_r$ rather than the random force field $ x \xi_\mu$. 
For a statistically uniform system ($x$-independent temperature and 
magnitude of the quenched disorder) with the Gaussian statistics of the  
disorder $\xi_r$, the evolution rate $V$ to the leading order is given 
by the following expression (compare with the thermally activated hopping 
rate from \cite{blatter,vinokur}) 
\begin{equation}
V\sim \exp\left(-D/\tau^2\right),
\label{creep}
\end{equation}
where $D$ is the variance of the random potential and $\tau$ is 
the temperature.   In the present context both local variance 
$D$ and the ``evolution temperature" $\tau$ depend on $x$, however we 
expect that this formula holds adiabatically if the typical 
waiting time $V^{-1}$ is large as compared to the relaxation time $V_0^{-1}$.  
The average evolution temperature $\langle\tau\rangle= \mu x/ 2 $, and 
variance is  $D\equiv V_0^2\langle \xi_r^2\rangle= V_0^2\Delta_r^2 /3$.
The evolution rate is thus determined by the following asymptotic
expression
\begin{equation}
V\sim V_0  \exp\left(-\frac{ 4 V_0^2}{3 \mu^2}\frac{\Delta_r^2}{x^2}\right)
= V_0 \exp\left(-\frac{N^2}{3}\frac{\Delta_r^2}{x^2}\right)
\label{creep1}
\end{equation}
(here we added the pre-exponential factor $V_0$ in order to match 
the results with the disorder-free case). According to Eq.(\ref{creep1}), 
the evolution rate $V$ grows monotonously with $x$ and finally 
approaches the disorder-free rate  $V_0$. 
Therefore, in a long-time limit the quenched potential $\xi_r$
becomes irrelevant because the growing with $x$ evolution temperature 
smears out the pinning potential and further promote the growth of $V$. 
However, for $x$ not too large ($x \ll N  \Delta_r$) one can distinguish
two different regimes of the (initial) evolution. If $\Delta_r \gg1$, 
in the absence of the thermal fluctuations a particle would be pinned by the 
random potential. Due to thermal fluctuations the particle will creep 
toward larger $x$, slowly increasing the velocity from $V=0$. 
In contrast, if $\Delta_r \sim O(1) $, i.e. above the depinning 
threshold, the particle starts to move with the finite velocity 
$V < V_0$, and after some time achieves asymptotic velocity $V_0$. 
Thus, the model exhibits ``thermal depinning" and ignores completely 
fitness fluctuations for large average fitness values.  Both regimes 
are seen in numerical simulations of the discrete model.  Shown in 
Fig. 3 are $\langle \langle x(t)  \rangle \rangle$  for 
different values of $\Delta_r$. We see that for  large $t$ all lines have
the same slope which coincides with the evolution rate on a smooth
landscape $V_0 =N\mu/2$ (see also inset in Fig. 3). Intermediate regime
exhibits a wide range of evolution rates which depend on $\Delta_r$.

In conclusion, we have shown that two types of quenched randomness  in the
fitness space have a different effect on the population evolution.  
Quenched randomness of the replication rate slows down the evolution only at
the initial stage since the increase of the  "evolution temperature" eventually
smooths out ruggedness and evolution proceeds at a rate corresponding to the 
smooth landscape case. However, in the long-time limit quenched disorder 
of the mutation rate being amplified by the increasing replication rate,  
dominate the evolution temperature growth.
In the limit $x, $ $ t \to \infty$  the evolution exhibits ultra-slow
logarithmic growth of averaged mean fitness $\langle \langle x \rangle\rangle$ .
Individual runs  are characterized by long intervals of almost constant mean 
fitness  $\langle x \rangle$ interrupted by spontaneous changes, a dynamics
which is usually interpreted as punctuated equilibrium. 
In this work, we have made simplest {\em ad hoc} 
assumptions regarding the statistical properties of the fitness landscape.
An important task for a future work will be to establish connections between
``standard'' multi-dimensional sequence space evolution models such as NK-model
and one-dimensional fitness space model described here.

We are grateful to David Kessler, Herbert Levine and Douglas Ridgway for 
useful discussions. This work was supported by the U.S. Department of 
Energy under contracts W-31-109-ENG-38 and DE-FG03-96ER14592.  The work 
of IA was also supported by  NSF-Office of Science and Technology  Center 
under contract No. DMR91-20000.

\vspace{-.7in}
\leftline{\psfig{figure=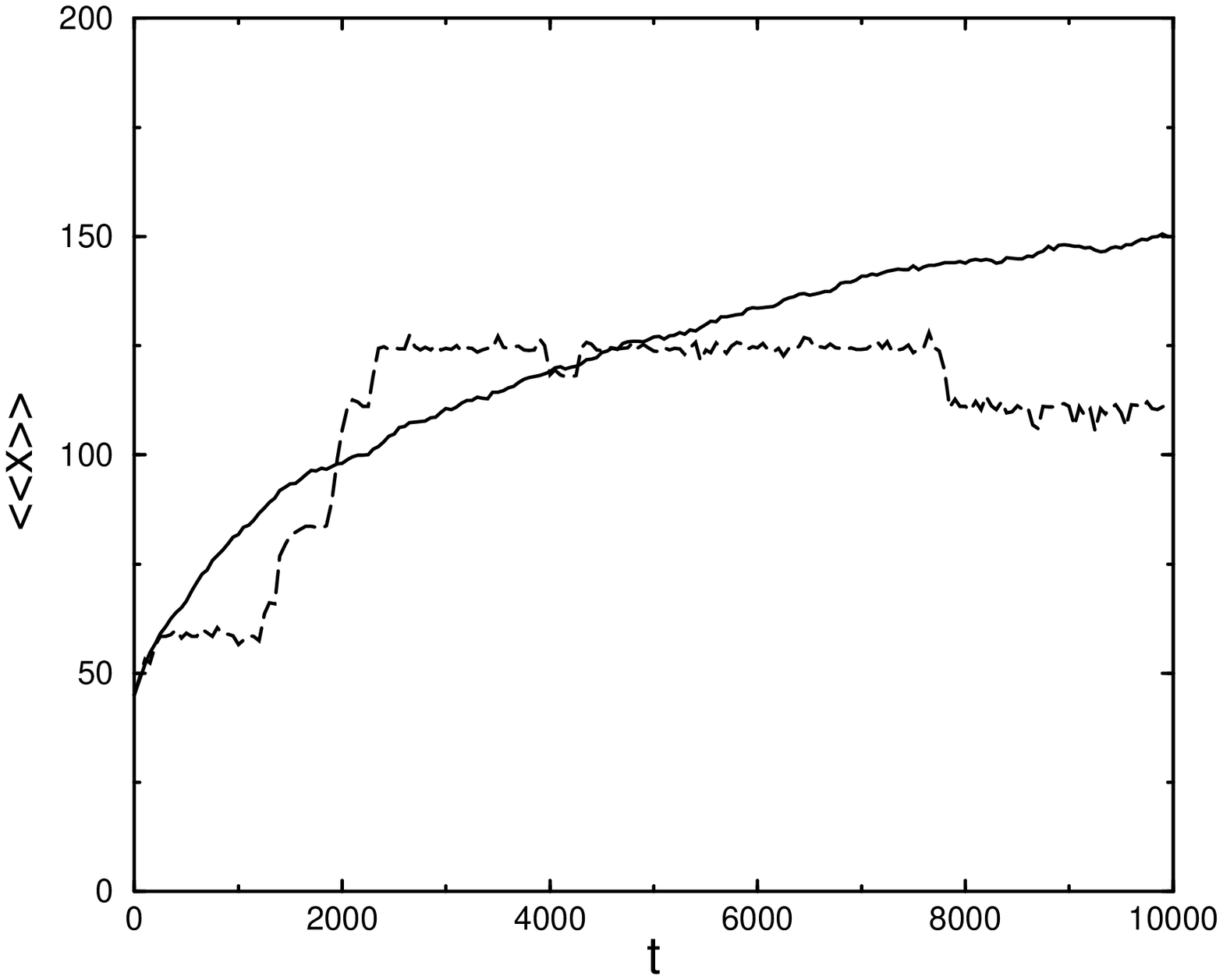,height=2.5in}}
\vspace{-.3in}
\begin{figure} 
\caption{
Averaged over 50 runs mean fitness $\langle \langle x(t)  \rangle \rangle$
for $N=50$ and $\mu=0.02$, $\Delta_\mu =0.7$, and $\Delta_r =0$ (solid line). 
Dashed line shows  $\langle x(t) \rangle$ for a single run.} 
\end{figure} 

\vspace{-.8in}
\leftline{\psfig{figure=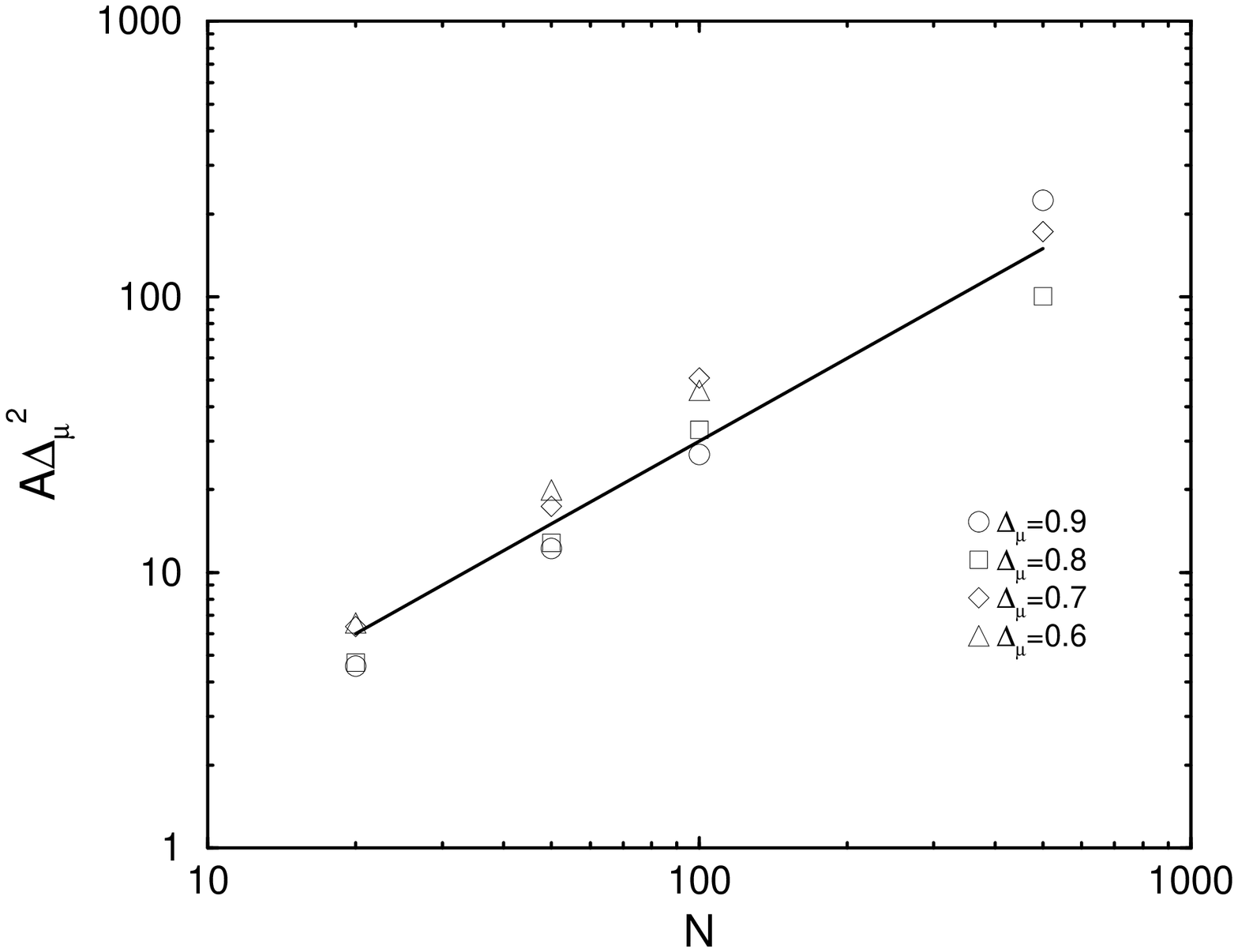,height=2.5in}}
\vspace{-.3in}
\begin{figure} 
\caption{Scaling of pre-logarithm factor $A$  with $N$ and $\Delta_\mu$, obtained
by fitting $\langle  \langle x(t) \rangle  \rangle=$ $ A\log(t+C_0)+x_0$.   
Solid line indicates a unit slope.
} 
\end{figure} 

\vspace{-.8in}
\leftline{\psfig{figure=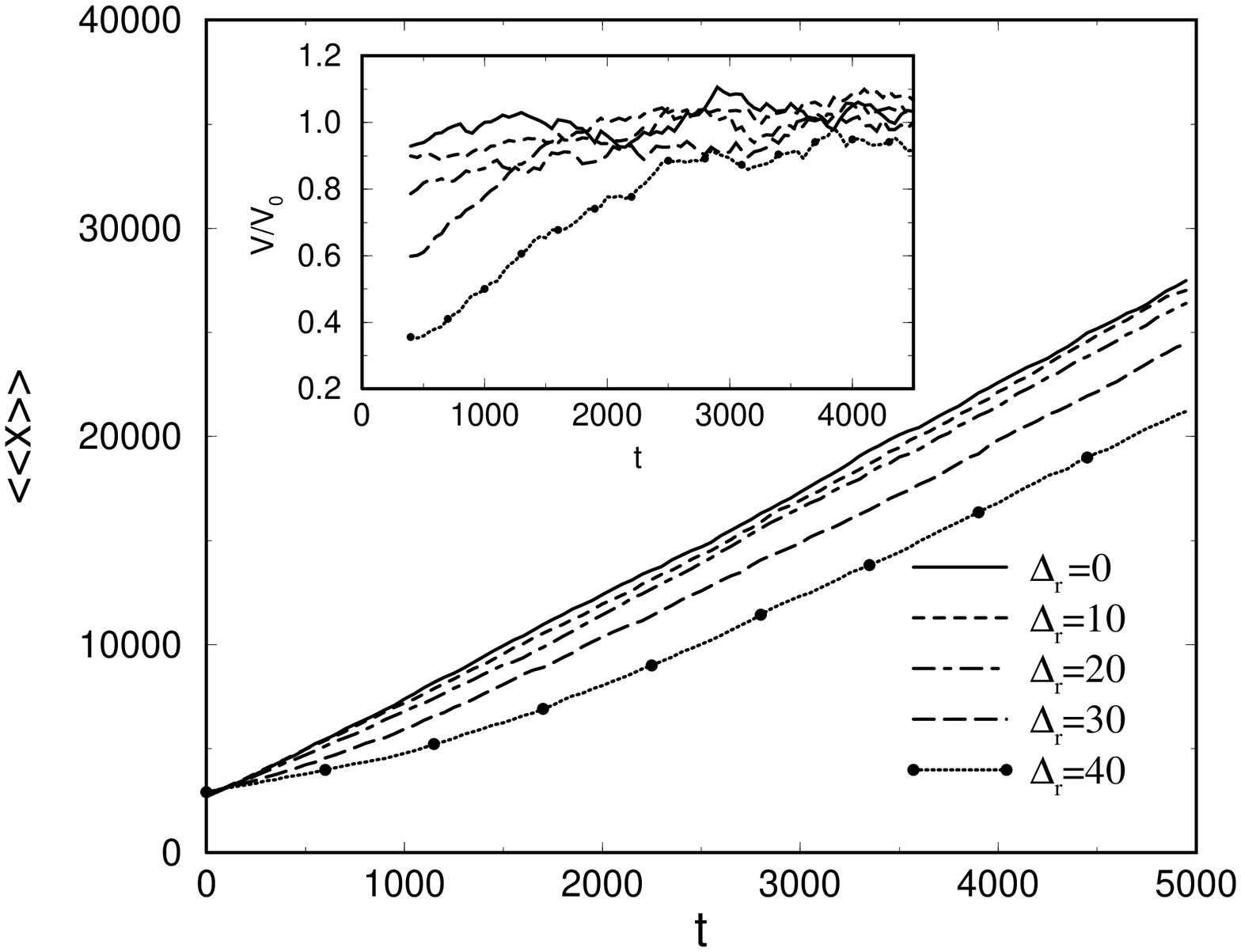,height=2.5in}}
\vspace{-.3in}
\begin{figure} 
\caption{Averaged over 100 runs mean fitness $ \langle \langle x(t) 
\rangle \rangle$ for $N=100$,  $\mu=0.1$, $\Delta_\mu =0$ and 
different $\Delta_r$. Inset shows $V = \partial_t \langle\langle x 
\rangle\rangle$. } 
\end{figure} 
\end{document}